\documentclass[aps,prl,twocolumn,showpacs,10pt]{revtex4}%
\usepackage{acronym}
\usepackage{amsfonts}
\usepackage{amsmath}
\usepackage{amssymb}
\usepackage{graphicx}%
\renewcommand{\mathbf}{\boldsymbol}

\begin{document}
\title{Relativistic treatment of spin-currents and spin-transfer torque}
\author{A. Vernes$^{1)}$, B. L. Gy\"{o}rffy$^{1,2)}$, and P. Weinberger$^{1)}$}
\affiliation{$^{1)}$ Center for Computational Materials Science, Technical University
Vienna, Gumpendorferstr. 1a, A-1060 Vienna, Austria $^{2)}$ H. H. Wills
Physics Laboratory, University of Bristol, Tyndall Avenue, Bristol BS8 1TL,
United Kingdom}
\date{\today}

\begin{abstract}
It is shown that a useful relativistic generalization of the conventional spin
density $\vec{s}(\vec{r},t)$ for the case of moving electrons is the
expectation value $(\vec{\mathcal{T}}(\vec{r},t),\,\mathcal{T}_{4}(\vec
{r},t))$ of the four-component Bargmann-Wigner polarization operator $T_{\mu
}=(\vec{T},T_{4})$ with respect to the four components of the wave function.
An exact equation of motion for this quantity is derived, using the
one-particle Dirac equation, and the relativistic analogues of the
non-relativistic concepts of spin-currents and spin-transfer torques are
identified. In the classical limit the time evolution of $(\vec{\tau},\tau
_{4})$, the integral of $(\vec{\mathcal{T}}(\vec{r},t),\,\mathcal{T}_{4}%
(\vec{r},t))$ over the volume of a wave packet, is governed by the equation of
motion first proposed by Bargmann, Michel and Telegdi generalized to the case
of inhomogeneous systems. In the non-relativistic limit it is found that the
spin-current has an intrinsic Hall contribution and to order $1\,\!/\,c^{2}$ a
spin-orbit coupling related torque appears in the equation of motion for
$\vec{s}(\vec{r},t)$. The relevance of these results to the theory of the
intrinsic spin Hall effect and current-induced switching are briefly discussed.
\end{abstract}


\maketitle

Central to the emerging technology of spin-based electronics, often
referred to as spintronics, is the observation that electron transport
can be influenced not only by coupling to the charge but also to the
spin of the current carrying electrons.~\cite{AFS02} A striking
example of charge-current being effected by the magnetic state of the
conductor is the giant magneto-resistance (GMR) phenomenon. Evidently,
the complementary effects of charge-currents inducing changes in the
magnetization of the conductors are also of interest. An example of
these is the current-induced switching first predicted by J. C.
Slonczewski and independently by L. Berger.~\cite{Slo96,Ber96} As is
now well established, it is due to the spin-transfer torque that a
spin-polarized current can exert on the magnetization of the structure
through which it flows.~\cite{SM06} However, the details of how such
torques come about have not been, as yet, fully explored. In
particular, all discussions of the problem are currently based on
non-relativistic quantum mechanics and hence neglect the spin-orbit
coupling. The purpose of this letter is to present a fully
relativistic theory of spin-currents and the above spin-transfer
torque in order to provide a conceptual framework in which the spin
and orbital degrees of freedoms can be treated on equal footing.

Another topic, to which such developments are relevant, is the spin Hall
effect intensively studied in semiconductor spintronics.~\cite{SMSC06,ERH06}
It involves a spin-current flowing perpendicularly to a charge-current in a
sample of finite width. Interestingly, it implies spin accumulation at the
edges and the possibility of spin-injection into an adjacent sample without
the presence of magnetic or exchange fields. Here, spin-orbit coupling is the
central issue and a source of difficulty is the lack of a well-defined
spin-current in a spin-orbit coupled system.~\cite{ERH06} It is hoped that the
polarization-current introduced in this letter will clarify this matter considerably.

A moving electron carries with it a spin and this moving spin amounts to a
spin-current. Classically, it is given by the tensor product,
$\overleftrightarrow{J}_{\mathrm{cl}}=\vec{s}_{\mathrm{cl}}\otimes\vec{v}$, of
the velocity vector $\vec{v}$ and a classical spin vector $\vec
{s}_{\mathrm{cl}}$. Quantum mechanically, for an electron described by a
two-component wave function $\phi=\left(
\begin{array}
[c]{c}%
\phi_{\uparrow}\\
\phi_{\downarrow}%
\end{array}
\right)  $, with spin components $\phi_{\uparrow}$ and $\phi_{\downarrow}$, it
is given by the tensor density%
\begin{equation}
\overleftrightarrow{J}_{\mathrm{s}}=\phi^{+}\,
\left(\mathbf{\vec{\sigma}}\otimes\vec{J}\right)\,\phi
\ ,\label{eq:spncurrnr}%
\end{equation}%
where $\vec{J}=i\hbar\, (\overleftarrow{\nabla} -
\overrightarrow{\nabla}) \,/\, (2m_{\mathrm{e}})$, the wave functions
and therefore also the spin-current
$\overleftrightarrow{J}_{\mathrm{s}}$ are evaluated at the space time
point $(\vec{r},\,t)$. The physical significance of
$\overleftrightarrow {J}_{\mathrm{s}}$ becomes apparent if we study
the time evolution of the spin density defined by
$\vec{s}=\phi^{+}\,\mathbf{\vec{\sigma}}\,\phi$. From the
time-dependent Schr\"{o}dinger equation which includes a Zeeman term
of the form $-\mu_{\mathrm{B}}\mathbf{\vec{\sigma}}\cdot\vec{B}$ we
find
\begin{equation}
\dfrac{d\vec{s}}{dt}+\nabla\cdot\overleftrightarrow{J}_{\mathrm{s}}=\dfrac
{e}{m_{\mathrm{e}}}\,\vec{s}\times\vec{B}\ .\label{eq:nreqmot}%
\end{equation}%
Clearly, $\nabla\cdot\overleftrightarrow{J}_{\mathrm{s}}$ may be
regarded as a torque which, in addition to the more familiar
microscopic Landau-Lifshitz torque $\vec{s}\times\vec{B}$, causes the
spin density $\vec{s}$ at the point $\overrightarrow{r}$ to evolve in
time. As explained, at length, in the insightful review of Stiles and
Miltat this spin-transfer torque depends linearly on the
charge-current and plays a central role in current-induced
switching.~\cite{SM06}

Note that for $\vec{B}=0$ Eq.~(\ref{eq:nreqmot}) is a continuity equation for
the spin density $\vec{s}$ and as such it follows via the Noether theorem from
the fact that $\vec{s}$ is a conserved quantity. The difficulty of
generalizing the continuity equation to spin-orbit coupled systems, such as
described by the Dirac equation or by various model Hamiltonians, e.g.\ , like
those used in semiconductor spintronics,~\cite{ERH06,PB02} arises from the
circumstance that in these cases the spin operator no longer commutes with the
Hamiltonian and hence the conventional spin density is not conserved as the
time evolves. In what follows this dilemma is resolved by the choice of a
convenient, covariant description of the spin-polarization, as an alternative
to that afforded by the usual spin operators in non-relativistic quantum
mechanics. In the following, for the sake of simplicity only the case of
non-interacting positive energy Dirac electrons will be considered.

For electrons described by the Dirac equation in the standard representation
it is common to refer to
\begin{equation}
\mathbf{\vec{\Sigma}}=\left(
\begin{array}
[c]{cc}%
\mathbf{\vec{\sigma}} & \mathbf{0}\\
\mathbf{0} & \mathbf{\vec{\sigma}}%
\end{array}
\right)
\label{eq:spinop}
\end{equation}
as the $4\times4$ Pauli spin operator. However, it corresponds to the spin of
an electron only in its rest frame and hence its use is not convenient in the
case of many moving electrons. Moreover, as mentioned above, it does not
commute even with the field-free Dirac Hamiltonian $\mathcal{H}_{\mathrm{D} }
= c \mathbf{\vec{\alpha}} \cdot\vec{p} + \mathbf{\beta}m_{\mathrm{e}}c^{2}$
and hence the corresponding density is not that of a conserved quantity. A
more suitable approach for describing the spin-polarization of moving
electrons is to use the four-component polarization operator $T_{\mu}%
\equiv(\vec{T},T_{4})$ introduced by Bargmann and Wigner.~\cite{BW48}
The salient features of this approach and its relations to other,
alternatives, are fully discussed in a comprehensive review article by
Fradkin and Good.~\cite{FG61} Here they will be merely referred to as
the need arises.

For the case of one electron, in the presence of an electromagnetic field
described by the vector potential $\vec{A}=\vec{A}(\vec{r},t)$ and a scalar
potential $V=V(\vec{r},t)$, the four-component polarization operator is
defined by
\begin{equation}
\left\{
\begin{array}
[c]{rcl}%
\vec{T} & = & \mathbf{\beta}\,\mathbf{\vec{\Sigma}}-i\Sigma_{4}\,\dfrac
{\mathbf{\vec{\Pi}}}{m_{\mathrm{e}}c}\\
&  & \\
T_{4} & = & i\mathbf{\,\vec{\Sigma}}\cdot\dfrac{\mathbf{\vec{\Pi}}%
}{m_{\mathrm{e}}c}%
\end{array}
\right.  \ ,\label{eq:polop4vct}%
\end{equation}
where the canonical momentum operator takes its usual form: $\mathbf{\vec{\Pi
}}=(\vec{p}-e\vec{A})\mathbf{I}_{4}$, with $\mathbf{I}_{4}$ being the
$4\times4$ unit matrix, $\ \mathbf{\vec{\Sigma}}$ is the spin operator defined
in Eq.~(\ref{eq:spinop}). For future reference note that\ $\mathbf{\ \vec
{\Sigma}}$ is part of the four-component operator $\Sigma_{\mu}\equiv
(\mathbf{\vec{\Sigma}},\,\Sigma_{4})$ whose 4th component is defined as
$\Sigma_{4}=-i\ \mathbf{\gamma}_{5}$, see e.g.\ Ref.~\onlinecite{Ros61}. It
is also of interest to note that both $T_{\mu}$ and $\Sigma_{\mu}$ are
covariant axial four-vectors.

From the point of view of our present concern the most important property of
$T_{\mu}$ is that commutes with the field-free Dirac Hamiltonian. Thus, as
will be shown below, the corresponding vector density satisfies a continuity
equation. To see the connection between the non-relativistic spin operator
$\mathbf{\vec{\sigma}}$ and $\vec{T}$ it is useful to note that the latter is
related to the magnetization of an electron in its rest frame by a Lorentz boost.

To derive a relativistic analogue of Eq.~(\ref{eq:nreqmot}) one has to
calculate the first derivative with respect to the time of the polarization
densities $\mathcal{\vec{T}}=\mathcal{\vec{T}}(\vec{r},t)$ and $\mathcal{T}%
_{4}=\mathcal{T}_{4}(\vec{r},t)$ defined by
\[
\mathcal{\vec{T}}=\psi^{+}\vec{T}\psi\qquad\mathrm{and}\qquad\mathcal{T}%
_{4}=\psi^{+}T_{4}\psi\ ,
\]
where $\psi^{+}=\psi^{+}(\vec{r},t)$ is the adjoint (conjugate transpose) of
the four-component solution $\psi=\psi(\vec{r},t)$ of the time-dependent Dirac
equation corresponding to the Hamiltonian $\mathcal{H}_{\mathrm{D}%
}=\mathcal{H}_{\mathrm{D}}(\vec{r},t)=c\mathbf{\vec{\alpha}}\cdot
\mathbf{\vec{\Pi}}+\mathbf{\beta}m_{\mathrm{e}}c^{2}+eV\mathbf{I}_{4}$. By
using the chain rule for all four components $\mu=1,\ldots,4$ in
Eq.~(\ref{eq:polop4vct}),
\[
\dfrac{d\mathcal{\vec{T}}_{\mu}}{dt}=\dfrac{\partial\psi^{+}}{\partial
t}T_{\mu}\psi+\psi^{+}\dfrac{\partial T_{\mu}}{\partial t}\psi+\psi^{+}T_{\mu
}\dfrac{\partial\psi}{\partial t}\ ,
\]
and the relations
\[
\dfrac{\partial\psi}{\partial t}=\dfrac{1}{i\hbar}\mathcal{H}_{\mathrm{D}}%
\psi\ ,\qquad\dfrac{\partial\psi^{+}}{\partial t}=-\dfrac{1}{i\hbar}\psi
^{+}\mathcal{H}_{\mathrm{D}}^{+}%
\]%
\[
\dfrac{\partial\vec{T}}{\partial t}=\mathbf{\gamma}_{5}\dfrac{e}%
{m_{\mathrm{e}}c}\dfrac{\partial\vec{A}}{\partial t}\ ,\qquad\dfrac{\partial
T_{4}}{\partial t}=-\dfrac{ie}{m_{\mathrm{e}}c}\mathbf{\ \vec{\Sigma}}%
\cdot\dfrac{\partial\vec{A}}{\partial t}\ ,
\]
after some lengthy but straightforward algebra one arrives at
\begin{equation}
\dfrac{d\mathcal{\vec{T}}}{dt}+\nabla\cdot\overleftrightarrow{\mathcal{J}%
}^{\,\prime}+\nabla\mathcal{J}^{\,\prime\prime}=\dfrac{e}{m_{\mathrm{e}}%
}\,\mathcal{\vec{S}}\times\vec{B}-\dfrac{ie}{m_{\mathrm{e}}c}\,\vec
{E}\,\mathcal{S}_{4}\ ,\label{eq:releqmot-a}%
\end{equation}
and
\begin{equation}
\dfrac{d\mathcal{T}_{4}}{dt}+\nabla\cdot\left(  \mathcal{\vec{J}}%
_{4}^{\,\,\,\prime}-\mathcal{\vec{J}}_{4}^{\,\,\prime\prime}\right)
=\dfrac{ie}{m_{\mathrm{e}}c}\,\mathcal{\vec{S}}\cdot\vec{E}%
\ ,\label{eq:releqmot-b}%
\end{equation}
where $\vec{B}=\vec{B}(\vec{r},t)$ is the magnetic induction vector and
$\vec{E}=\vec{E}(\vec{r},t)$ the electric field intensity. In
Eqs.~(\ref{eq:releqmot-a}) and (\ref{eq:releqmot-b}), the four-component
density $\mathcal{S}_{\mu}\equiv(\mathcal{\vec{S}},\,\mathcal{S}_{4})$ is
given by
\[
\mathcal{\vec{S}}=\psi^{+}\mathbf{\vec{\Sigma}\,}\psi,\qquad\mathcal{S}%
_{4}=\psi^{+}\Sigma_{4}\psi
\]
and the polarization-current density tensors are defined as
\begin{equation}%
\begin{array}
[c]{rlllrll}%
\overleftrightarrow{\mathcal{J}}^{\,\prime} & = & c\,\psi^{+}\left(  \vec
{T}\otimes\mathbf{\vec{\alpha}}\right)  \psi &  & \mathcal{\vec{J}}%
_{4}^{\,\,\,\prime} & = & c\,\psi^{+}\left(  T_{4}\mathbf{\vec{\alpha}%
}\right)  \psi\\
&  &  &  &  &  & \\
\mathcal{J}^{\,\prime\prime} & = & \psi^{+}\left(  2c\,\,\mathbf{\beta\gamma
}_{5}\right)  \psi &  & \mathcal{\vec{J}}_{4}^{\,\,\,\prime\prime} & = &
\psi^{+}\left(  2\,\dfrac{\vec{\pi}}{m_{\mathrm{e}}}\times\mathbf{\vec{\alpha
}}\right)  \psi
\end{array}
\ ,\label{eq:relcurr}%
\end{equation}
such that
\[
\nabla\cdot\overleftrightarrow{\mathcal{J}}^{\,\prime}
=\displaystyle\sum\limits_{j=\mathrm{x}\,,\mathrm{y}\,,\mathrm{z}}
\partial_{j}\left[ \psi^{+}\left( \vec{T}\mathbf{\alpha}_{j}\right)
  \psi\right] \ .
\]

Although the above relations bear some resemblance to Eq.~(\ref{eq:nreqmot})
the problem of generalizing it to relativistic quantum mechanics is not yet
completed because Eqs.~(\ref{eq:releqmot-a}) and (\ref{eq:releqmot-b} ) are
not a closed set of equations for the polarization density in terms of
\ corresponding currents. To proceed further one must derive the equation of
motion for the four-component auxiliary density $\mathcal{S}_{\mu}$. Following
the same route as in the case of $\mathcal{T}_{\mu}\equiv(\mathcal{\vec{T}%
},\,\mathcal{T}_{4})$ one straightforwardly finds that
\begin{equation}
\dfrac{d\mathcal{\vec{S}}}{dt}-ic\,\nabla\mathcal{S}_{4}=\dfrac{m_{\mathrm{e}%
}c}{\hbar}\mathcal{\vec{J}}_{4}^{\,\,\prime\prime}+i\,\nabla\times
\mathcal{\vec{J}}\label{eq:releqmot-c}%
\end{equation}
and
\begin{equation}
i\,\dfrac{d\mathcal{S}_{4}}{dt}-c\nabla\cdot\mathcal{\vec{S}}=i\,\dfrac
{m_{\mathrm{e}}c}{\hbar}\mathcal{J}^{\,\prime\prime}\ ,\label{eq:releqmot-d}%
\end{equation}
with $\mathcal{\vec{J}} = c\,\psi^{+}\mathbf{\vec{\alpha}}\psi$ being
the relativistic probability current density. Remarkably, some of the
same currents which appear in the equations for $\mathcal{T}_{\mu}$
determine $\mathcal{S}_{\mu}$ and hence for a given set of currents in
Eq.~(\ref{eq:relcurr}), Eqs.~(\ref{eq:releqmot-a}),
(\ref{eq:releqmot-b}), (\ref{eq:releqmot-c}) and (\ref{eq:releqmot-d})
can be solved for $\mathcal{T}_{\mu}$ and $\mathcal{S}_{\mu}$.

The relations in Eqs.~(\ref{eq:releqmot-a}) - (\ref{eq:releqmot-d}) are the
central result of this letter. Namely, the comparison of
Eqs.~(\ref{eq:nreqmot}) and (\ref{eq:releqmot-a}), without electromagnetic
fields, uniquely identifies the polarization-current density as
$\overleftrightarrow{\mathcal{J}}^{\,\prime}+\mathcal{J}^{\,\prime\prime
}\mathbf{I}_{3}$ and its divergence as the relativistic generalization of the
conventional spin-current density and spin-transfer torque, respectively.
Indeed, in the case of a vanishing electromagnetic field
Eq.~(\ref{eq:releqmot-a}) reduces to a continuity equation for the
polarization density in the same manner as Eq.~(\ref{eq:nreqmot}) is a
continuity equation for the magnetization density.

To shed light on the physical content of these results they will now be
examined in two separate limits. First the classical, $\hbar\rightarrow0$,
then the non-relativistic, $c^{-2}\rightarrow0$ limit will be studied.

The aim of the classical limit is to find a dynamical description of
the polarization of an electron whose orbital motion is classically
given by the position vector $\vec{r}_{\mathrm{cl}}(t)$ as prescribed
by a relativistic classical mechanical equation of motion. This is the
case of interest in Ref.~\onlinecite{FG61} and phenomenologically is
treated in Ref.~\onlinecite{LL99b}. In short one assumes that
$\psi(\vec{r},t)$ describes a wave packet centered at the position
vector $\vec{r}_{\mathrm{cl}}(t)$ and moving with a velocity
$\vec{v}_{\mathrm{cl}}$. The size of the wave packet is to be taken to
be large compared with the Compton wave length $\hbar\,/\,mc$ but very
much smaller then the scale on which the external electromagnetic
field vary. Moreover, it contains momentum components only near
$m_{\mathrm{e}} \vec{v}_{\mathrm{cl}}$. In the lights of these
assumptions it is natural to define a four-vector $\tau_{\mu}=$\ 
$(\vec{\tau},\tau_{4})$ as the integral of the density
$\mathcal{T}_{\mu}$ over the variable $\vec{r}$ within a volume
$\Omega$ as
\[
\vec{\tau}\equiv\vec{\tau}(t)=\displaystyle\int\limits_{\Omega}d^{3}%
r\,\mathcal{\vec{T}}(\vec{r},t)\ ,\quad\tau_{4}\equiv\tau_{4}%
(t)=\displaystyle\int\limits_{\Omega}d^{3}r\,\mathcal{T}_{4}(\vec{r},t)
\]
and derive an equation of motion for $\tau_{\mu}(t)$ from those for
$\mathcal{T}_{\mu}$ as given in Eqs.~(\ref{eq:releqmot-a}) and
(\ref{eq:releqmot-b}). Evidently, the property $\tau_{\mu}(t)$ is to be
associated with a classical particle described by $\vec{r}_{\mathrm{cl}}(t)$
and $\vec{v}(t)$. Following the procedure of Ref.~\onlinecite{FG61} we
find
\begin{align}
\frac{d\vec{\tau}}{dt}+\left.  \displaystyle\int\limits_{\Omega}d^{3}%
r\,\nabla\cdot\overleftrightarrow{\mathcal{J}}^{\,\prime}(\vec{r}%
,t)\right\vert _{\mathrm{cl}} &  =\bar{\gamma}^{-1}\,\vec{\tau}\times\vec
{B}(\vec{r}_{\mathrm{cl}},t)\label{eq:cleqmot}\\
&  -i\frac{e}{m_{\mathrm{e}}c}\bar{\gamma}^{-1}\,\tau_{4}\vec{E}(\vec
{r}_{\mathrm{cl}},t)\ ,\nonumber
\end{align}
where $\bar{\gamma}$ is the Lorentz factor $\left(  1-v^{2}/c^{2}\right)
^{-1\,/\,2}$ and , as indicated, the external fields are evaluated at the
current position $\vec{r}_{\mathrm{cl}}(t)$ of the particle. Noting the result
$\left\langle \mathcal{S}_{\mu}\right\rangle \simeq$ $\ \bar{\gamma}%
^{-1}\,\tau_{\mu}$ of Ref.~\onlinecite{FG61}, where $\left\langle
\mathcal{S}_{\mu}\right\rangle $ denotes the classical limit of the density
$\mathcal{S}_{\mu}$ integrated over $\Omega$, the form of the right-hand side
of the above relation readily follows from Eq.~(\ref{eq:releqmot-a}). The
term involving $\mathcal{J}^{\prime\prime}$ was also shown
to be zero by following the procedure of Ref.~\onlinecite{FG61}.

Before commenting on its most interesting feature, namely the torque
on the left-hand side, it is reassuring to note that without it
Eq.~(\ref{eq:cleqmot}) is exactly that of Bargmann, Michel and Telegdi
(BMT)~\cite{BMT59} discussed at length by Landau.~\cite{LL99b} The
extra factor of $\bar{\gamma}$ is due to the fact that the derivative
on the left-hand side is with respect to the global time $t$ and not
the proper time in the rest frame of the electron as in
Ref.~\onlinecite{LL99b}. To clarify the physical content of the BMT
equation, Landau introduces a classical spin polarization vector in
the rest frame, here denoted by $\vec{s}(t)$, and demands that when it
is Lorentz boosted into the global frame, $\vec{s}(t)$ would be equal
to $\vec{\tau}(t)$ in this letter. This allows him to convert the
equation for $\vec{\tau}(t)$ to one describing $\vec{s}(t)$. In the
interest of brevity his result is recalled here only to order
$c^{-2}$:
\[
\frac{d\vec{s}}{dt}\simeq\frac{e}{m_{\mathrm{e}}c}\,\vec{s}\times\vec{B}%
+\frac{e}{2m_{\mathrm{e}}c}\,\vec{s}\times\left(  \vec{E}\times\dfrac{\vec{v}%
}{c}\right)  \ .
\]
Clearly, the second term on the right-hand side is a\ torque -- due to an
effective magnetic induction vector $\vec{E}\times\vec{v}\,/\,c$\ -- a moving
electron experiences in an electric field $\vec{E}$. In other words it is the
simplest manifestation of spin-orbit coupling. For example, a Rashba
Hamiltonian analogue of this term is the main mechanism of the intrinsic spin
Hall effect in the work of Sinova et al.~\cite{SCN+04}

The novel and the most interesting feature of Eq.~(\ref{eq:cleqmot}) is the
torque on the left-hand side. For a uniform $\overleftrightarrow{\mathcal{J}%
}^{\,\prime}$, e.g.\ , in a bulk material, $\nabla\cdot\overleftrightarrow
{\mathcal{J}}^{\,\prime}$ is zero. But when the effective classical particle
crosses an interface or a domain wall with spin-dependent properties, the
volume integral can be converted into an integral over a closed surface. If
this surface includes the interface, whose differential oriented surface is
$d\vec{A}$, then the torque is $(\overleftrightarrow{\mathcal{J}}^{\,\prime
+}-\overleftrightarrow{\mathcal{J}}^{\,\prime\_}$ $)\cdot d\vec{A}$, where
$\pm$ refers to the opposite sites of the interface. Thus, just as in the
non-relativistic case, see Ref.~\onlinecite{SMSC06}, this torque is due to an
excess in the spin-current flowing in and out of the surface region causing a
polarization accumulation. This then supports in very explicit terms, the
identification of $\overleftrightarrow{\mathcal{J}}^{\,\prime}+\mathcal{J}%
^{\,\prime\prime}\mathbf{I}_{3}$ as the fully relativistic generalization of
the non-relativistic definition of spin-current. Notably, in the classical
limit one might approximate $\overleftrightarrow{\mathcal{J}}^{\,\prime}%
\simeq\vec{\tau}\otimes\vec{u}$, where $\vec{u}$ is the relativistic velocity
vector\ $\bar{\gamma}\vec{v}$ and $\vec{\tau}$ is described by the BMT
equation (\ref{eq:cleqmot}). This is a rather satisfactory result given the
form of the non-relativistic classical formula $\overleftrightarrow
{\mathcal{J}}^{\,\prime}\simeq\vec{s}\otimes\vec{v}$ quoted at the beginning
of this letter.

Finally, in what follows we examine the lowest order corrections to the
non-relativistic theory. Working to the order $1\,/\,c$, after considerable
algebra we find that the equations for $\mathcal{\vec{T}}$ and $\mathcal{\vec
{S}}$ satisfy one and the same equation of motion, and the
polarization-current is given by
\begin{align}
\overleftrightarrow{\mathcal{J}}^{\left(  \mathrm{1}\right)  }  &
= \phi^{+}\,
\left(\mathbf{\vec{\sigma}}\otimes\vec{J} \right)\,\phi 
  - \left(  \phi^{+}\mathbf{\vec{\sigma }}\phi\right)  
\otimes\dfrac{e\vec{A}}{m_{\mathrm{e}}} \nonumber\\
& + \frac{\hbar}{2m_{\mathrm{e}}}\,
\phi^{+}\, \left\{\mathbf{\vec{\sigma}}\otimes\left[
\left(\overleftarrow{\nabla} + \overrightarrow{\nabla}\right) 
\times\mathbf{\vec{\sigma}}\right]\right\}\,\phi \ . 
\label{eq:spncurrfirst}
\end{align}
Evidently, the first two terms in Eq.~(\ref{eq:spncurrfirst}) are the
generalization of the conventional
$\overleftrightarrow{J}_{\mathrm{s}}$ in Eq.~(\ref{eq:spncurrnr}) to
the case of a non-vanishing vector potential.  The third term
$\delta\!  \overleftrightarrow{\mathcal{J}}^{\left( \mathrm{1}\right)
}$ is a consequence of the internal contribution to the probability
current density due to the moving dipole moment $\delta\!
\vec{J}_{\mathrm{int}} = \hbar\, \nabla \times \left(\phi^{+}
  \mathbf{\vec{\sigma}}\phi\right) \,/\,
(2m_{\mathrm{e}})$,~\cite{LL99b} and its form readily follows from the
anzatz $\delta\! \overleftrightarrow{\mathcal{J}} ^{\left(
    \mathrm{1}\right) } = \phi^{+} ( \mathbf{\vec{{\sigma}}} \otimes
\,\delta\!  \hat{\vec{J}}_{\mathrm{int}}) \phi$.~\cite{JLZ06}
Moreover, just as $\delta\! \vec{J}_{\mathrm{int}}$ does not
contribute to the divergence in the continuity equation for the
probability density,
$\nabla\cdot\delta\!\overleftrightarrow{\mathcal{J}}^{\left(
    \mathrm{1}\right) }$ is identically zero and gives rise to no
torque.

To the order of $1\,/\,c^{2}$ there are many more contributions. These will be
discussed in a separate publication. Here only one term $\delta
\overleftrightarrow{\mathcal{J}}_{\mathrm{SOC}}^{\left(  \mathrm{2}\right)  }%
$, which is clearly to be associated with the spin-orbit coupling, is
highlighted:%
\begin{align}
\delta\overleftrightarrow{\mathcal{J}}_{\mathrm{SOC}}^{\left(  \mathrm{2}%
\right)  }  & =\dfrac{ie\hbar}{2m_{\mathrm{e}}^{2}c^{2}}\,\vec{E}\cdot\left(
\tilde{\phi}^{+}\mathbf{\vec{\sigma}}\tilde{\phi}\right)
\,\mathbf{I}_{3} \label{eq:spncurrsoc}\\
& +\dfrac{e\hbar}{2m_{\mathrm{e}}^{2}c^{2}}\,\left(
\begin{array}
[c]{ccc}%
0 & +E_{\mathrm{z}} & -E_{\mathrm{y}}\\
-E_{\mathrm{z}} & 0 & +E_{\mathrm{x}}\\
+E_{\mathrm{y}} & -E_{\mathrm{x}} & 0
\end{array}
\right)  \left(  \tilde{\phi}^{+}\tilde{\phi}\right)  \ ,\nonumber
\end{align}
where $\tilde{\phi}$ is $\phi$ renormalized as in Ref.~\onlinecite{LL99b}.

Remarkably, the off-diagonal terms have the form required by the spin
Hall effect.~\cite{ERH06} It means that for example an electric field,
and presumably a charge-current, along the $\mathrm{z}$ axis, a spin
polarization along the $\mathrm{x}$ axis implies a
polarization-current in the $\mathrm{y}$ direction. Interestingly,
this term is the only contribution obtained, if one uses, in a simple
minded derivation, the anomalous velocity,~\cite{CB01b}
$\vec{v}_{\mathrm{a}} = -e\hbar\,( \mathbf{\vec{\sigma}} \times
\vec{E} )\,/\,(4m_{\mathrm{e}}^{2}c^{2})$, and the non-relativistic
definition of the spin-current density $\tilde{\phi}^{+}
[\mathbf{\vec{\sigma}} \otimes \vec{v} + (\vec{v} \otimes
\mathbf{\vec{\sigma}}) ^{\mathrm{T}} ]
\tilde{\phi}$.~\cite{ERH06,JLZ06} Thus the first term in
Eq.~(\ref{eq:spncurrsoc}) is a nontrivial consequence of our more
general, fully relativistic treatment of the polarization.  Clearly,
it implies that for a charge current along the electric field there
will be a helicity dependent contribution to the spin-current.

Whilst the above reference to the spin Hall effect can not be taken as
an explanation for the observed spin Hall currents due to the
smallness of the vacuum coupling constant $\lambda_{\mathrm{SOC}} =
e\hbar\,/\,(2m_{\mathrm{e}}^{2}c^{2})$, see Ref.~\onlinecite{ERH06},
the presence of a relevant term in Eq.~(\ref{eq:spncurrsoc}) suggests
that an intrinsic spin Hall effect is a generic feature of spin-orbit
coupled systems and therefore of the relativistic quantum mechanics.

In conclusion it should be stressed that the above discussion was
confined to a one-electron theory based on a one-electron Dirac
equation. Nevertheless, the results establish the line of reasoning a
relativistic generalization of the corresponding many-particle theory
has to take. In particular it will lead to a relativistic version of
the semi-classical transport theory for the current-induced switching
dynamics,~\cite{SM06} or for that of the spin Hall
effect.~\cite{CSS+04} It will also facilitate the corresponding
generalization of the time-dependent density functional theory of
Capelle et al.,~\cite{CVG01} and will readily provide a framework for
first-principles calculation using fully relativistic methods such as
the screened RKKR.~\cite{ZHSW05} Interestingly, it will also enter the
relativistic Fermi liquid theory of Baym and Chin.~\cite{BC76}

Financial support by the Vienna Science and Technology Fund (WWTF), the
Wolfgang-Pauli Institut (WPI) and the Vienna Institute of Technology (TU
Vienna) is gratefully acknowledged.

\end{document}